\documentclass[%
 reprint,
 amsmath,amssymb,
 aps,
prb,
dvipdfmx
]{revtex4-1}
\usepackage[dvipdfmx]{graphicx}
\usepackage[dvipdfmx]{color}
\usepackage{dcolumn}

\usepackage{amsmath} \usepackage{amssymb}
\usepackage{accents} \usepackage{bbm}
\usepackage{mathrsfs}
\usepackage{accents}

\newcommand{\bs}[1]{\boldsymbol{#1}} 
\newcommand{\tr}[1]{\mathrm{Tr}\left[#1\right]}

\newcommand{\ut}[1]{\undertilde{#1}} 
\newcommand{\ve}[0]{\varepsilon}

\definecolor{dgreen}{RGB}{00, 120, 00}
\usepackage{hyperref}
\hypersetup{colorlinks=true, linkcolor=blue, urlcolor=blue, citecolor=blue, }
\usepackage{booktabs}

\usepackage{ulem}

\begin{document}
\preprint{APS/123-QED}

\title{Effects of the phase coherence on the local density of states in
superconducting proximity structures}

\author{Shu-Ichiro Suzuki$^{1,2}$}
\author{Alexander A. Golubov$^{2,3}$}
\author{Yasuhiro Asano$^{3,4,5}$}
\author{Yukio Tanaka$^{1}$}

\affiliation{$^{1}$Department of Applied Physics, 
Nagoya University, Nagoya 464-8603, Japan}
\affiliation{$^{2}$MESA$^+$ Institute for Nanotechnology, 
University of Twente, 7500 AE Enschede, The Netherlands}
\affiliation{$^{3}$Moscow Institute of Physics and Technology, 
141700 Dolgoprudny, Russia} 
\affiliation{$^{4}$Department of Applied Physics, 
Hokkaido University, Sapporo 060-8628, Japan}
\affiliation{$^{5}$Center of Topological Science and Technology, 
Hokkaido University, Sapporo 060-8628, Japan}

\date{\today}

\begin{abstract}

We theoretically study the local density of states in superconducting
proximity structure where two superconducting terminals are attached
to a side surface of a normal-metal wire. Using the quasiclassical
Green's function method, the energy spectrum	is obtained for both of
spin-singlet $s$-wave and spin-triplet $p$-wave junctions.  In both of
the cases, the decay length of the proximity effect at the zero
temperature is limited by a depairing effect due to inelastic
scatterings.  In addition to the depairing effect, in $p$-wave
junctions, the decay length depends sensitively on the transparency at
the junction interfaces, which is a unique property to odd-parity
superconductors where the anomalous proximity effect occurs.

\end{abstract}

\pacs{???}
\maketitle


\section{\label{sec:introduction} Introduction}

\begin{figure}[b!]
	\centering
  \includegraphics[width=0.40\textwidth]{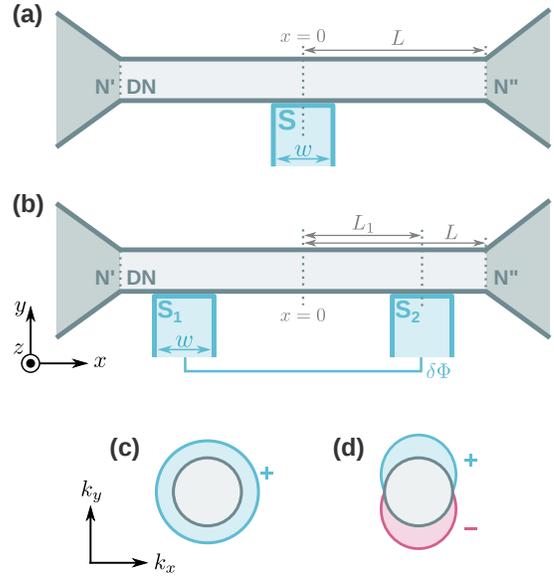}
	\caption{Schematics of (a) T-shaped and (b) Volkov-Takayanagi (VT)
	junctions.  The N'/DN interfaces are located at $x=\pm L$. The
	barrier potential is present only at the S/DN interfaces. 
	The widths and the thickness of the
	wires are assumed much narrower and thinner than the coherence
	length. The superconductor(s) is attached to
	the DN at $x=0$ in (a) and at $x=\pm L_1$ (b). The
	superconductors have the phase difference $\delta \Phi$ in (b). 
	Schematics of
	the (c) $s$\,-wave and (d) $p$\,-wave pair potentials in momentum
	space. The inner
	circles indicate the Fermi surface. The sign means
	the phase of the pair potential. }
	\label{fig:Sche}
\end{figure}

The proximity effect is a phenomenon observed in a normal metal (N)
attached to a superconductor (SC)\cite{Deutscher}.  Cooper pairs
penetrating into an N causes superconducting-like phenomena such as
the screening of magnetic fields and the suppression of the local
density of states (LDOS) at the Fermi level (zero energy).  The
penetration length of Cooper pairs is limited by the thermal coherence
length $\xi_T=\sqrt{\mathscr{D}/2\pi T}$, where $\mathscr{D}$ is the
diffusion constant in the N and $T$ is the temperature.  Indeed, the
Josephson current is present only when the spacing between two SCs
$L_1$ is shorter than $\xi_T$.\cite{DeGennes_1964} 
Although $\xi_T$ is the typical length scale of the proximity effect,
Volkov and Takayanagi have shown that the characteristic length
depends on observables\cite{Volkov_PRL_1996, Volkov_PRB_1997}.  They
studied the conductance of a normal-metal wire whose side surface is
connected to two superconducting terminals [See
Fig.~\ref{fig:Sche}(b).].  The conductance depends on the phase
difference of the two SCs even when $L_1 \gg
\xi_T$\cite{Volkov_PRL_1996, Volkov_PRB_1997}.  Thus this phenomenon
is named the long-range phase-coherent effect.

The analysis by Volkov and Takayanagi is unfortunately restricted to
the weak-proximity-effect regime, 
where the solutions of the linearized Usadel equation describe 
the long-range phase-coherent effect.
However, the magnitude of the proximity effect is generally sensitive to 
the transparency of an N/SC 
interface and the pairing symmetry of the superconductor.
The strong proximity effect leads a gap-like energy spectrum at 
low energy in the LDOS
\cite{Volkov_Physica_1993, Golubov_JLT_1988, Golubov_JETP_1989, Golubov_PRB_1997}.
The boundary condition for the quasiclassical Green's function\cite{Kuprianov_ZETF_1988,
Nazarov_PRL_1994,Nazarov_Superlattice_1999} enables these analysis.

Taking the essence of the circuit
theory\cite{Nazarov_PRL_1994,Nazarov_Superlattice_1999} into account,
a boundary condition for the quasiclassical
Usadel Green's function 
at an N/SC interface has been
derived\cite{Tanaka_PRB_2004,Tanaka_PRB_2005,Tanaka_PRL_2003,Proximityd2}. This boundary condition
enables to describe junctions of unconventional SCs 
such as high-$T_c$ cuprate, spin-triplet SCs, and topological SCs. 
It has been well established that
the Andreev bound states (ABSs) due to the unconventional pairing\cite{TK95}
modifies the proximity effect in various ways.
In an N/$d$\,-wave junction, 
the proximity effect can not contribute to ensemble-averaged values
over random-impurity 
configurations\cite{Tanaka_PRL_2003,Proximityd2}. However, 
the amplitude of the Josephson current in each $d$\,-wave/N/$d$\,-wave
junction can exceed the ensemble-averaged Josephson current for
the $s$\,-wave/N/$s$\,-wave junctions\cite{Asano_PRB_2001,Asano_PRB_2006}. 
Spin-triplet parings\cite{shivaram_PRL_1986,
Choi_PRL_1991, Maeno_Nature_1994, Rice_1995, Ishida_Nature_1998,
Graf_PRB_2000, saxena_Nature_2000, aoki_Nature_2001,
Mackenzie_RMP_2003, Huy_PRL_2007,Kashiwaya11, YMachida_PRL_2012}
cause several anomalies (i.e., the anomalous proximity effect) such as large zero-energy peaks 
in the LDOS in the N\cite{Tanaka_PRB_2004, Tanaka_PRB_2005,Tanaka_PRBR_2005} 
and resonant charge transport through a dirty N
\cite{Tanaka_PRB_2004, Tanaka_PRB_2005,
Tanaka_PRBR_2005, Asano_PRL_2006, Asano2013, Ikegaya_PRB_2015,
Ikegaya_PRB_2016}. The anomalous proximity effect is a result of
the penetration of the ABSs\cite{ABS,ABSb,Hu,TK95,ABSR1} 
into the normal metal or equivalently the
appearance of odd-frequency Cooper pairs in the normal metal\cite{Tanaka_PRL_2007,
Tanaka_PRL_2007_1,Tanaka_PRB_2007,Eschrig_JLTP_2007,Asano_PRL_2007,
Asano_PRL_2011, Asano2013, Higashitani_PRL_2013, SIS1, SIS2, 
Ikegaya_JOP_2016, SIS3}. 
Such unusual phenomena have attracted much attention these days because 
they are equivalent to the physics of Majorana Fermions appearing 
topologically nontrivial SCs\cite{Read, Ivanov01,
qi11, Fu_PRL_2008, Tanaka_PRL_09, Fu_PRL_2009, Akhmerov_PRL_2009,
JDSau_PRB_2010, Tanaka_review_2012, Mizushima_JP_2015,
Sato_review_2016, Mizushima_JPSJ_2016, Sato_review_2017, NW-exp2015,
NW-exp2016,lutchyn_review_2017, Majo_exp_PRX2017, Majo_exp_Nat2017,
Majo_exp_gul_2018, Majo_Sestoft_2017, Majo_chen2016, SIS4_PRB_2018,
LABOldeOlthof_PRB_2018}.  
At present, however, we have never known how the anomalous proximity 
effect modifies the long-range phase-coherent phenomena.



In this paper, we study the local density of state (LDOS) in a
wire of a diffusive normal metal (DN) by solving numerically
the quasiclassical Usadel equation in the regime of the 
strong proximity effect. 
We consider two types of proximity structures: 
T-shaped junction shown in Fig.~\ref{fig:Sche}(a) and
Volkov-Takayanagi (VT) junction shown in Fig.~\ref{fig:Sche}(b).
We found in the T-shaped junction that the quasiparticle density of states 
depends strongly
on the barrier potential at the junction interface. 
In the VT junction, the LDOS between the
two superconducting electrodes depends sensitively on the phase difference 
of the two superconducting electrode. 
In an in-phase junction, the LDOS in DN between the $s$\,-wave
($p$\,-wave) superconducting electrodes shows the zero-energy dip
(peak), whereas such dip and peak structures vanish in an out-of-phase
junction because of the destructive interference of Cooper
pairs.  In an $s$-wave junction, the phase-coherent effect
is spatially limited by a decay length due to depairing of Cooper
pairs. In a $p$-wave junction, in addition to the depairing effect,
the low transparency at the junction interface limits the
long-range phase-coherent effect as well. 

This paper is organized as following.  In Sec.~\ref{sec:KelUsa}, the
Keldysh-Usadel formalism and the system we consider are explained.  In
Sec.~\ref{sec:T}, we discuss the calculated LDOS for the T-shaped
junction. In Sec.~\ref{sec:VT}, we shown the LDOS in the
Volkov-Takayanagi junction and discuss the long-range coherence. 
In particular, we focus
on the junction-length and depairing-ratio dependences of the LDOS. 
We summarize this study in Sec.~\ref{sec:con}. 


\section{Keldysh-Usadel formalism}
\label{sec:KelUsa}

\subsection{Usadel equation}
In this paper, we consider the junctions of a diffusive normal metal
(DN) where superconducting (S) wires are attached to a side surface of the
DN as shown in Fig.~\ref{fig:Sche}.  We refer to the junction shown in
Figs.~\ref{fig:Sche}(a) and \ref{fig:Sche}(b) as T-shaped and
Volkov-Takayanagi (VT) junctions, respectively. 
In the T-shaped junction, a narrow S wire is
attached to a wire of the DN at $|x| < w/2$ 
and $y=0$ with a finite interface resistance $R_b$, where $w$ is the
width of the S arm which is much shorter than the
superconducting coherence length in the diffusive system 
$\xi_0= \sqrt{\mathscr{D}/2\pi T_c}$. In the VT junction, narrow
S wires are attached at $|x \mp L_1| < w/2$.  The DN is connected to
lead wires of clean normal metal at $x = \pm L$, but sufficiently narrow	and thin in the $y$ and
$z$ directions (i.e., $L_{y(z)} \ll \xi_0$). 

The Green's function in the DN obeys the Usadel equation\cite{Usadel}: 
\begin{align}
  & \mathscr{D} \bs{\nabla}
	\left( \mathbb{G} \bs{\nabla} \mathbb{G} \right)
	+ i \left[ \mathbb{H}, \mathbb{G} \right]_-
	= 0, 
	\\[2mm]
	& \mathbb{G}(\bs{r} ,\ve) = \left( \begin{array}{cc}
	 \check{g}^R(\bs{r} ,\ve)& \check{g}^K(\bs{r},\ve) \\[1mm]
	 0                       & \check{g}^A(\bs{r},\ve) \\
	\end{array} \right), 
\end{align}
where $\mathscr{D}$ is the diffusion constant in the DN, $\check{g}^X$ with
$X=K$, $R$, and $A$ are the Keyldysh, retarded, and advanced
components of the Usadel Green's function, and $\mathbb{H} =
\mathrm{diag}[\check{H}^R, \check{H}^A]$. 
Assuming the width of the DN is much narrower than $\xi_0$,
we can ignore the spatial variation of the
Green's function in the $y$ direction in the DN. Namely, one need to
consider a one-dimensional diffusive system where the Usadel equation
is reduced to 
\begin{align}
  & \mathscr{D} \partial_x 
	\left( \mathbb{G} \partial_x \mathbb{G} \right)
	+ i \left[ \mathbb{H}, \mathbb{G} \right]_-
	+ \mathbb{S} \Theta_S(x)
	= 0, 
  \label{eq:Usadel-ori}
\end{align}
where the last term $\mathbb{S}(x,\ve)$ represents effects of the
S wires. 
The source term $\mathbb{S}(x,\ve)$ is reduced from 
the boundary condition in the $y$ direction\cite{Volkov_PRL_1996, Volkov_PRB_1997}. 
The step-like function is unity only at
the place where the S wires are attached: 
$\Theta_S(x) = \Theta(w/2-|x    |)$ for the T-shaped junction and 
$\Theta_S(x) = \Theta(w/2-|x-L_1|) + \Theta(w/2-|x+L_1|)$ 
for the VT junction.
In this paper, the symbols written
in a bold mean matrices in the Keldysh space, and the accents
$\check{\cdot}$ and $\hat{\cdot}$ means matrices in particle-hole
space and spin space. 
The identity matrices in particle-hole and spin space 
are respectively denoted by $\check{\tau}_0$ and $\hat{\sigma}_0$. 
The Pauli matrices are denoted by $\check{\tau}_j$ and $\hat{\sigma}_j$ 
with $j \in [1,3]$. 
The Keldysh-Usadel equation is supplemented by the so-called
normalization condition: $\mathbb{G} \mathbb{G} = \mathbbm{1}$. 
The Keldysh Green's function can be obtained from the following
relation: 
\begin{align}
  & \check{g}^K = \check{g}^R \check{F} -  \check{F}\check{g}^A, \\[1mm]
  & \check{F}   = \check{\tau}_0 f_L + \check{\tau}_3 f_T, 
\end{align}
where $f_L = \tanh(\ve/2T)$ and  $f_T=f_T(x,\ve)$ describes the
derivation from equilibrium. 

The LDOS is related to the retarded and advanced
components of the Usadel Green's function. The Usadel equation for
$X=R$ and $A$ in one
dimension is given by 
\begin{align}
  \mathscr{D} \partial_x
	\left( \check{g}^X  \partial_x \check{g}^X \right)
	+ i \left[ \check{H}^X, \check{g}^X \right]_-
	+ \check{S}^X \Theta_S(x)
	= 0, 
  \label{eq:Usadel-RA}
	\\[2mm]
	\check{g}^X(x,\ve) = \left( \begin{array}{rr}
	     \hat{g} ^X &  \hat{f}^X \\[1mm]
	-\ut{\hat{f}}^X & -\hat{g}^X \\[1mm]
	\end{array} \right). 
\end{align}
where $\check{H}^X = \tilde{\ve}^X \check{\tau}_3$. The factor
$\bar{\ve}^X$ depends on $X$: 
$\bar{\ve}^R = \ve + i \gamma$ and 
$\bar{\ve}^A = \ve - i \gamma$, 
where $\ve$ and
$\gamma$ being the energy and the depairing ratio due to inelastic
scatterings. 
In this paper, we assumed that there is no spin-dependent
potential, that the Cooper pairs has one single spin component
(i.e., $\hat{\Delta} = \Delta_\mu i \hat{\sigma}_\mu \hat{\sigma}_2$
with $\Delta_\mu$ being the scalar pair potential), and 
the phase difference between two SCs is $\delta \Phi = 0$
or $\pi$ (i.e., no-current states).
In this case, one can parametrize the matrix structure of the Green's
functions as follows: 
\begin{align}
  & \hat{g} ^X = \hat{\sigma}_0 g^X,  \\[1mm]
  & \hat{f} ^X = f^X_\mu (i\hat{\sigma}_\mu \hat{\sigma}_2), 
	  \hspace{6mm}
  \ut{\hat{f}}^X = f^X_\mu (i \hat{\sigma}_2 \hat{\sigma}_\mu), 
\end{align}
where $\mu$ is related to the direction of the synthetic spin of Cooper
pairs: $\mu=0$ and $\mu=1$-$3$ correspond to the spin-singlet and
spin-triplet parings. The Usadel equation can be simplified by this
parametrization:
\begin{align}
  \mathscr{D} \partial_x
	\left( \tilde{g}^X  \partial_x \tilde{g}^X \right)
	+ i \left[ \ve^X\tilde{\tau}_3, \tilde{g}^X \right]_-
	+ \tilde{S}^X \Theta_S(x)
	= 0, 
  \label{eq:Usadel-RA}
	\\[2mm]
	\hspace{6mm}
	\tilde{g}^X(x,\ve) = \left( \begin{array}{rr}
	   g^X &  f^X \\[1mm]
	  -f^X & -g^X \\[1mm]
	\end{array} \right), 
\end{align}
where we have introduced the symbol $\tilde{\cdot}$ meaning a $2
\times 2$ matrix in spin-reduced particle-hole space [e.g., $ \check{g}^X(x,\ve) =
\tilde{g}^X(x,\ve) \otimes \hat{\sigma}_0$]. Here we assumed the phase
difference between two SCs is $0$ or $\pi$ which
simplifies the relation between $f^X$ and $\ut{f}^X$ as discussed in
Appendix. 

The standard angular parametrization makes the Usadel equation much
simpler\cite{Volkov_Physica_1993, Golubov_JLT_1988, Golubov_PRB_1997}.
The Green's function can be well parametrized by the following
parameterization:
\begin{align}
  \tilde{g}^X 
	&=
    \tilde{\tau}_3 \cosh \theta
  +i\tilde{\tau}_2 \sinh \theta, \\[1mm]
	&= \left[ \begin{array}{rr}
	   \cosh \theta &  \sinh \theta \\[1mm]
	 - \sinh \theta & -\cosh \theta \\
	\end{array} \right], 
\end{align}
where we omit the index $X$ from $\theta
= \theta^X(x,\ve)$.  This parametrization always satisfies the
normalization condition:
$\tilde{g}^X \tilde{g}^X = \tilde{\tau}_0$. 
The Usadel equation is reduced by this parametrization: 
\begin{align}
  & \mathscr{D} \frac{\partial^2 \theta}{\partial {x}^2}
	+ 2 i \bar{\ve} \sinh \theta 
	+ \Theta_S(x) {S}(x,\ve) = 0. 
  \label{eq:Usadel-th} 
\end{align}

\subsection{Effects of superconducting terminals}

The last term in the left hand side of Eq.~\eqref{eq:Usadel-th} [i.e.,
$S(x,\ve)$] represents the effect of the S arms attached to the side
surface of the DN \cite{Volkov_Physica_1993,Golubov_PRB_1997}. The
typical boundary conditions\cite{Kuprianov_ZETF_1988} are no longer
available for junctions of unconventional SCs. In order to discuss the
proximity effect by unconventional pairings, one must employ the
so-called Tanaka-Nazarov
condition\cite{Tanaka_PRB_2004,Tanaka_PRB_2005}, which is an extension
of the circuit theory\cite{Nazarov_PRL_1994,
Nazarov_Superlattice_1999}.  The source term ${S}$ is derived from the
boundary condition in the $y$ direction. We employ the Tanaka-Nazarov
boundary condition discussed in
Refs.~[\onlinecite{Tanaka_PRL_2003,Proximityd2,Tanaka_PRB_2004,Tanaka_PRB_2005}]:
\begin{align} & \left. \frac{d \theta}{d y} \right|_{y=0} =
\gamma_B^{-1} \langle F \rangle_\phi, \label{eq:Tanaka-Nazarov}
\\[2mm]
	&~F = \frac{ -2 T_N (f_S \cosh \theta_0 - g_S \sinh \theta_0)}
	{(2-T_N)\Xi    +T_N (g_S \cosh \theta_0 - f_S \sinh \theta_0)},
	\end{align} where $\gamma_B = R_b / R_N \xi_0$ is the barrier
	parameter with $R_b$ and $R_N$ being the interface resistance per
	unit area and the specific resistance of the DN, $T_N(\phi) =
	\cos^2\phi/ (\cos^2\phi + z_0^2)$ is the transmission coefficient of
	an N/N interface with a barrier potential $\hbar v_F z_0$, $\phi$ is
	the angle of the momentum measured from the $k_y$ axis, and
	$\theta_0(x) = \theta(x)|_{y=0}$. The angle $\phi$ is measured from
	the $y$-axis. The angular bracket means angle average: 
$ \langle \cdots \rangle_\phi
	\equiv 
	\big( \int_{-\pi/2}^{\pi/2} \cdots \cos \phi~d\phi \big)
	\big( \int_{-\pi/2}^{\pi/2} T_N    \cos \phi~d\phi \big)^{-1}. 
$
The functions $g_S$ and $f_S$ can
be obtained from the Green's functions in a homogeneous ballistic
superconductor:
\begin{align}
  & g_S = g_{S+} + g_{S-}, \\[1mm]
  & f_S = 
	\left\{ \begin{array}{ll}
		f_{S+} + f_{S-}                  & \text{~for singlet SCs, } \\[1mm]
	  i(g_{S-} f_{S+} - g_{S+} f_{S-}) & \text{~for triplet SCs, } \\
	\end{array} \right. \\[2mm]
	& g_{S\pm}(\phi) = \frac{\ve}{\sqrt{ \ve^2 - |\Delta_\pm|^2}}, 
  ~~ f_{S\pm}(\phi) = \frac{\Delta_\pm}{\sqrt{ \ve^2 -
	|\Delta_\pm|^2}}, \\
	& \Xi = 1 + g_{S+} g_{S-} - f_{S+} f_{S-} 
\end{align}
where $\Delta_+(\phi) = \Delta_-(\pi-\phi)$. The pair potential
depends on the pairing symmetry of the superconductor: 
\begin{align}
  \Delta_+(\phi) = 
	\left\{ \begin{array}{ll}
		\Delta_0                            & \text{~for an $s$\,-wave, } \\[1mm]
	  \Delta_0 \cos [\phi-\alpha_\phi]    & \text{~for a $p$\,-wave,}\\[1mm]
	\end{array} \right. 
\end{align}
where $\Delta_0 \in \mathbb{R}$ is the amplitude of the pair potential
in a homogeneous superconductor and $\alpha_\phi$ parameterizes the
direction of the anisotropic superconductor. 
The boundary condition \eqref{eq:Tanaka-Nazarov} is transformed into 
the source term in the present case. The source term is given by 
\begin{align}
  S(x,\ve) = \gamma_B^{-1} \langle F(x,\ve,\phi) \rangle_\phi. 
\end{align}

The diffusivity changes the symmetry of Cooper pairs because only the
isotropic $s$\,-wave pairs can survive in diffusive systems. 
In the present case, the symmetry of S wires determines 
The symmetry of the Cooper pairs induced in the DN. 
In the $s$\,-wave junction, spin-singlet $s$-wave
Cooper pairs are induced, whereas spin-triplet
$s$\,-wave Cooper pairs are induced in the $p$\,-wave
junction\cite{Tanaka_PRL_2007,Tanaka_PRL_2007_1}. In order to satisfy the
Fermi-Dirac statistics, the spin-triplet Cooper pairs must belong to the
odd-frequency pairing symmetry\cite{Berezinskii}. 


\subsection{Boundary conditions}
The Usadel equation \eqref{eq:Usadel-th} is supplemented by the
boundary conditions. 
The boundary conditions for the T-shaped junction and the VT junction
without a phase difference are given by 
\begin{align}
          \theta(x,\ve)      \bigg|_{x=\pm L} = 0, \hspace{8mm}
  \frac{d \theta(x,\ve)}{dx} \bigg|_{x=0    } = 0. 
\end{align}
The boundary conditions for the VT junction
with the $\pi$-phase difference is given by 
\begin{align}
  \theta(x,\ve) \bigg|_{x=\pm L} = 0, \hspace{8mm}
  \theta(x,\ve) \bigg|_{x=0    } = 0. 
\end{align}
The details are written in Appendix. 

The LDOS $\nu(x,\ve)$ 
can be obtained from the Green's function: 
\begin{align}
  \nu(x,\ve) = \frac{\nu_0}{8}
	\tr{\check{\tau}_3 \left( \check{g}^R - \check{g}^A
	\right)}. 
\end{align}
where $\nu_0$ is the density of states per spin at the Fermi level 
in the normal states. 
In proximity structures, it is convenient to introduce 
the deviation of the LDOS: 
\begin{align}
  \delta \nu(x,\ve) =
  \frac{ \nu (x,\ve)-\nu_0 } {\nu_0}. 
	\label{eq:dos}
\end{align}

We solve numerically Eq.~\eqref{eq:Usadel-th} using the so-called
``forward elimination, backward substitution method''. 


\begin{figure}[tb]
	\centering
  \includegraphics[width=0.48\textwidth]{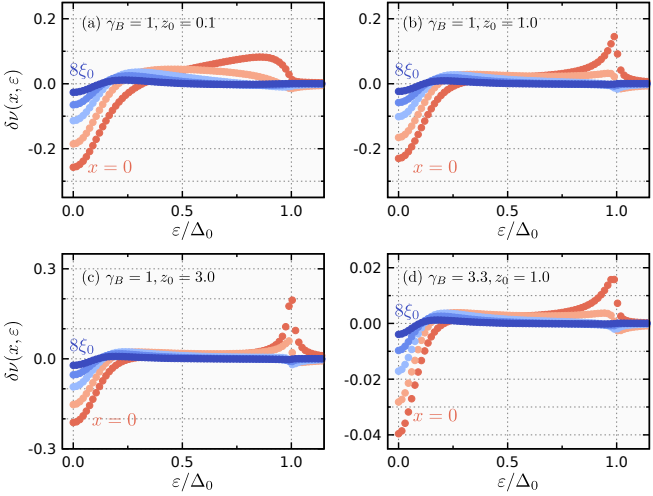}
	\caption{Deviations of the densities of states
	$\delta\nu(x,\ve)$ in the T-shaped junction with an $s$\,-wave
	superconducting wire. The results are obtained at $x=0$, $2\xi_0$,
	$4\xi_0$, $6\xi_0$, $8\xi_0$. 
	The barrier parameter is set to $\gamma_B = 1$ in (a), (b), and (c),
	$\gamma_B = 3.33$ in (d). The interface-potential parameter is set to 
  $z_0 = 0.1$ in (a), 
  $      1.0$ in (b) and (d), and 
  $      3.0$ in (c). 
	The length of the DN is set to $L=4\xi_0$. A superconductor with a
	width $w=0.3\xi_0$ is attached to the DN at $x=0$. The deparing
	ratio is set to $\gamma = 0.01\Delta_0$. 
	The structures such as coherence peak and low-energy dip become
	sharper with increasing $z_0$. The amplitude becomes smaller with
	increasing $\gamma_B$. }
	\label{fig:T_LDOS_TN_s}
\end{figure}

\section{T-shaped junctions}
\label{sec:T}
We first discuss the roles of the important interface parameters
(i.e., $z_0$ and $\gamma_B$) 
in a junction where a SC is attached to a side surface of
the DN.  The deviation of the LDOS $\delta \nu(x,\ve)$, which is given
in Eq.~\eqref{eq:dos}, in the T-shaped junction with an $s$\,-wave
SC are shown in Fig.~\ref{fig:T_LDOS_TN_s}.  The deviation
$\delta \nu$ is obtained at $x=0$ (beneath the S wire), $2\xi_0$,
$4\xi_0$, $6\xi_0$, $8\xi_0$.  The length of the DN and the width of
the S arm is set to $L=10\xi_0$ and $w=0.3 \xi_0$, respectively.  The
barrier parameter is set to $\gamma_B = 1$ in (a), (b), and (c),
$\gamma_B = 3.33$ in (d).  The interface-potential parameter is set to 
$z_0 = 0.1$ in (a), 
$      1.0$ in (b) and (d), and 
$      3.0$ in (c). 

In an $s$\,-wave junction, the coherence peak appears beneath the
S arm at the energy $\ve \sim \Delta_0$ because of the proximity effect\cite{Golubov_PRB_1997}. 
Simultaneously, at the
low energy, an energy dip appears reflecting the energy gap in the
S arm \footnote{We have confirmed that the anomalous Green's function
becomes pure imaginary, meaning that the conventional even-frequency Cooper pairs
are injected from the SC to the diffusive normal metal.}. The peak height and dip depth monotonically
decrease with increasing the distance from the S terminal. 

Comparing 
Figs.~\ref{fig:T_LDOS_TN_s}(a), \ref{fig:T_LDOS_TN_s}(b), and 
\ref{fig:T_LDOS_TN_s}(c), we can see that the coherence peak around
$\ve = \Delta_0$ becomes sharper and higher as $z_0$ increases. On the other
hand, the dip width in the energy and real space does not strongly 
depends on $z_0$. The dip width and depth are mainly determined by the 
spacing between normal lead wires (i.e., $2L$). We have confirmed that
the low-energy dip becomes narrower with increasing system
size\cite{Golubov_PRB_1997}. Comparing Figs.~\ref{fig:T_LDOS_TN_s}(d) with
\ref{fig:T_LDOS_TN_s}(b), we can see that the amplitude of $\delta
\nu$ becomes smaller with increasing the interface resistance (i.e.,
increasing of $\gamma_B$).

\begin{figure}[tbp]
	\centering
  \includegraphics[width=0.48\textwidth]{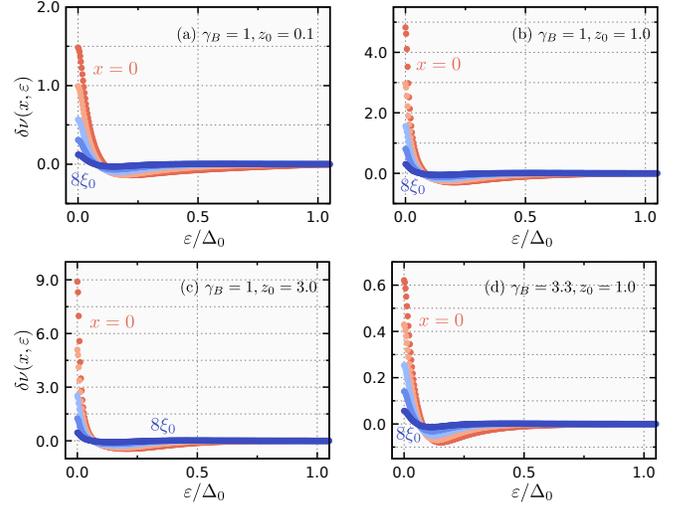}
	\caption{Deviations of the densities of states $\delta
	\nu(x,\ve)$ in the T-shaped junction with a $p$\,-wave SC. The
	parameters are set to the same values as those used in
	Fig.~\ref{fig:T_LDOS_TN_s}: $L=10\xi_0$, $w=0.3 \xi_0$, and $\gamma
	= 0.01\Delta_0$. The zero-energy peak appears because of the
	$p$\,-wave nature. The zero-energy peak becomes narrower and higher
	with increasing the interface potential $z_0$. }
	\label{fig:T_LDOS_TN_p}
\end{figure}
Contrary to the $s$\,-wave case, in the T-shaped junction with a
$p$\,-wave SC, the so-called zero-energy peak appears 
due to the anomalous proximity effect by odd-frequency spin-triplet
$s$\,-wave Cooper pairs 
\cite{Tanaka_PRB_2004,Tanaka_PRB_2005,Tanaka_PRL_2007,Asano_PRL_2007_2}
where topologically protected zero-energy states 
penetrate into the DN \cite{Ikegaya_PRB_2015,Ikegaya_PRB_2016}. 
\footnote{We have confirmed that the anomalous Green's function
becomes a real function, meaning that the odd-frequency Cooper pairs
are induced. }
Differing from the $d$-wave case (not shown), the zero
energy peak can survive in a $p$\,-wave junction even in a diffusive
system reflecting the orbital symmetry of odd-frequency pairing
\cite{Tanaka_PRL_2007} and the topological nature of a $p$\,-wave SC
\cite{Asano_PRB_2015,SIS2,Ikegaya_PRB_2016,SIS3}. 
The peak becomes higher but narrower in energy space with increasing
$z_0$. The peak width in real space, on the other
hand, does not strongly depend on the $z_0$. 
As happened in the $s$\,-wave junctions, $\gamma_B$ changes basically
only the amplitude of the deviation $|\delta \nu|$. 
The coherence peak around $\ve = \Delta_0$ does
not appear in the $p$\,-wave case. The zero-energy anomaly in 
$p$\,-wave T-shaped junctions can be observed by the charge transport
measurements\cite{Asano_PRL_2007_2}.

\section{Volkov-Takayanagi junctions}
\label{sec:VT}

\subsection{quasiparticle spectrum}

\begin{figure}[tb]
	\centering
  \includegraphics[width=0.40\textwidth]{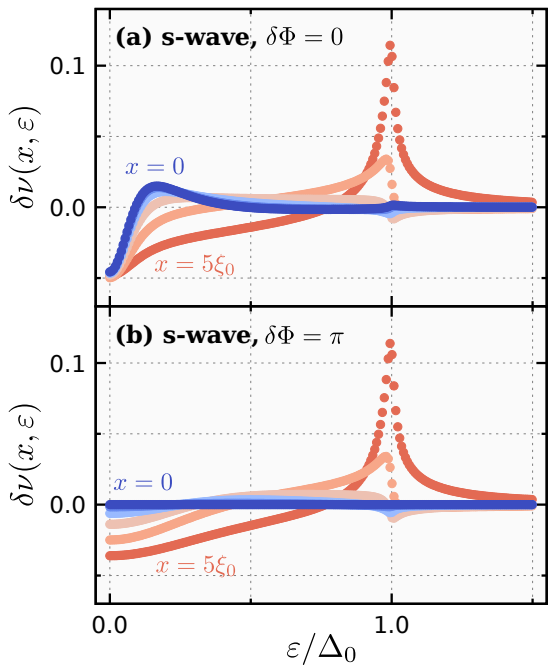}
	\caption{Correction of the local density of states (LDOS) in the VT junction
	with $s$\,-wave superconducting arms. The phase difference is set to
	(a) $\delta \Phi = 0$ and (b) $\delta \Phi = \pi$. The results are obtained 
	between the center of the junction (i.e., $x=0$) and the point where a
	superconducting arm is attached (i.e., $x=L_1$). The parameters are set to
	$L=6\xi_0$, $L_1 = 5\xi_0$, $w=0.3\xi_0$, $\gamma = 0.01\Delta_0$, 
	$\gamma_B = 1$, and $z_0 = 1$. 
	The LDOS at the center of the junction is modified when $\delta \Phi
	= 0$, whereas the correction vanishes 
	$\delta \Phi = \pi$. The results mean Cooper pairs from
	each superconductor interfere in the DN. }
	\label{fig:ldos_VT_s}
\end{figure}

\begin{figure}[tb]
	\centering
  \includegraphics[width=0.40\textwidth]{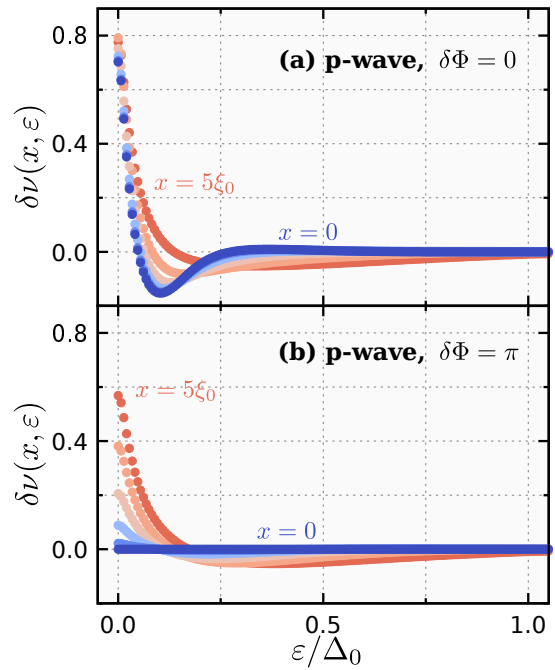}
	\caption{LDOS in the VT junction
	with $p$\,-wave superconducting arms. The results are plotted in the
	same manner as used in Fig.~\ref{fig:ldos_VT_s}. The parameters are
	set to the same values as those used in Fig.~\ref{fig:ldos_VT_s}. 
	The zero-energy peak spreads
	spatially between the two superconducting arms when $\delta \Phi
	=0$, whereas it vanishes when $\delta \Phi = \pi$ because of the
	long-range phase coherence. }
	\label{fig:ldos_VT_p}
\end{figure}

In a two-superconductor system such as Josephson junctions, the phase
difference between the two S wires affects significantly
on the quasiparticle spectrum in the junction. 
The LDOS in the VT junction with $s$\,-wave
SCs are shown in
Fig.~\ref{fig:ldos_VT_s}(a) and \ref{fig:ldos_VT_s}(b), where the
phase difference is set to $\delta \Phi= 0$ and $\pi$, respectively.
The parameters are set to $L=6\xi_0$, $L_1 = 5\xi_0$, $w=0.3\xi_0$,
$\gamma = 0.01\Delta_0$, $\gamma_B = 1$, and $z_0 = 1$.
When there is no phase
difference, there is an energy dip whose size is about $0.2\Delta_0$
at the zero energy. 
This energy dip spreads between the S arms even though
the spacing between the two arms is set to $2L_1 = 10 \xi_0$.

When the phase difference is $\delta \Phi = \pi$ the LDOS
at the center of the junction becomes completely flat as shown in
Fig.~\ref{fig:ldos_VT_s}(b). In addition, even at intermediate points,
the kink around $0.2\Delta_0$, which exists when $\delta \Phi
=0$, vanishes and $\delta \nu$ is more insensitive to $\ve$. As a result, the energy dip is no longer prominent in
Fig.~\ref{fig:ldos_VT_s}(b).  
These behavior can be interpreted in terms of the destructive
interference of Cooper pairs injected from the S arms.  The phase of
the anomalous Green's function describing the Cooper pairs is related
to the sign of the pair potential.  In the $\delta \Phi = \pi$
junction, the Cooper pairs from each arm have an opposite phase. In
other words, the pair amplitude of 
Cooper pairs cancel perfectly each other at the
center of a junction.  As a consequence, the LDOS at the
center becomes completely flat.  Reflecting this behavior, the Green's
function has an additional symmetry in real space $f^X(x,\ve) = -
f^X(x, \ve)$ (See Appendix for details). 

The LDOS in the $p$\,-wave VT junction are shown in
Fig.~\ref{fig:ldos_VT_p}(a) and \ref{fig:ldos_VT_p}(b), 
where the phase difference is set to $\delta
\Phi= 0$ and $\pi$, respectively. When $\delta \Phi = 0$, the
zero-energy peak spreads between the two S wires (i.e.,
$|x| \leq L_1$). The peak is the highest beneath the S 
wires and the lowest at the center of the junction. 
The low-energy dip at the center of the junction is more
prominent than that beneath the S wire. 
The dip width at $x=0$ is about $0.2 \Delta_0$ which is
comparable with that for the $s$-wave case shown in 
Fig.~\ref{fig:ldos_VT_s}(a). 
When $\delta \Phi = \pi$, as happened in the $s$-wave VT junction, 
the LDOS is completely flat at $x=0$. Moreover, 
the height of the zero energy peak is lower
than the $\delta \Phi = 0$ case due to the destructive interference of
the Cooper pairs injected from each SC.  

\begin{figure}[tb]
	\centering
  \includegraphics[width=0.40\textwidth]{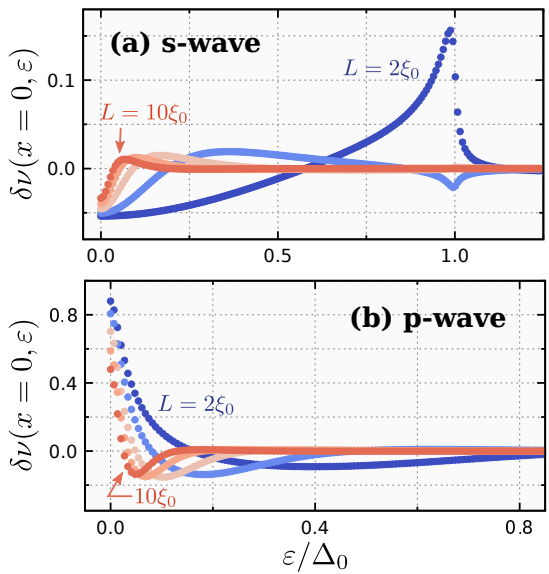}
	\caption{Junction-length dependence of the LDOS	at the center of VT junctions with (a) $s$-wave and (b) $p$\,-wave
	superconducting arms. The junction length is changed from $L=2\xi_0$
	to $10\xi_0$, where $\delta \Phi = 0$ and the interval between the
	superconducting wire and the normal lead is fixed at $L-L_1 = \xi_0$. 
	The other parameters are set to 
	$w=0.3\xi_0$, $\gamma = 0.01\Delta_0$, $\gamma_B = 1$, and $z_0 = 1$. }
	\label{fig:ldos-center}
\end{figure}

Differing from the typical $p$\,-wave Josephson
junction\cite{Asano_PRL_2006}, in the VT junction, the most
constructive and destructive interferences occur when $\delta \Phi =
0$ and $\delta \Phi = \pi$, respectively.  As shown in
Fig.~\ref{fig:Sche}(b), the S wires are attached to the
side surface which is normal to the $y$ axis. On the contrary, in the
typical Josephson junction, $p_x$\,-wave SCs attached in the $x$
direction. In the $p$\,-wave VT junction without a
phase difference, the anomalous Green's functions injected from both
of the S wires have the same sign. When the phase
difference is $\pi$, however, Cooper pairs from each S 
wire have opposite phase, which leads the destructive interference.

\subsection{Junction-length dependence}

The coherence is diminished with increasing the junction length. The
junction-length dependence of the LDOS at the
center of the VT junction with the $s$\,- and $p$\,-wave SCs are plotted
in Figs.~\ref{fig:ldos-center}(a) and \ref{fig:ldos-center}(b),
respectively. In the calculations, we set the phase difference $\delta
\Phi = 0$, $z_0=1$, and $L-L_1 = \xi_0$. In the $s$\,-wave VT junction, the
LDOS shows a dip structure at low energy even in a
sufficiently long junction. 
This energy dip becomes wider with decreasing
the junction length.
The height of the coherence peak 
around $\ve = \Delta_0$ strongly depends on the
junction length. With decreasing the
junction length, $\delta \nu |_{\ve=\Delta}$ is
almost the unity for $L>4\xi_0$, is negative for $L=4\xi_0$, and
becomes positive for $L=2\xi_0$.  In the short-junction limit, $\delta
\nu$ becomes qualitatively the same as that in the T-shaped
junction. 

The coherence in a $p$\,-wave junction modifies the LDOS 
as happened in the $s$\,-wave case. As shown in
Fig.~\ref{fig:ldos-center}(b), the zero-energy peak and the energy dip
can be seen even when $L=10\xi_0$. The width of the zero-energy peak
in energy space decreases monotonically with increasing the
junction length. The peak height at $x=0$ and $\ve=0$ decreases
monotonically with increasing the junction length. 

\begin{figure}[tb]
	\centering
  \includegraphics[width=0.48\textwidth]{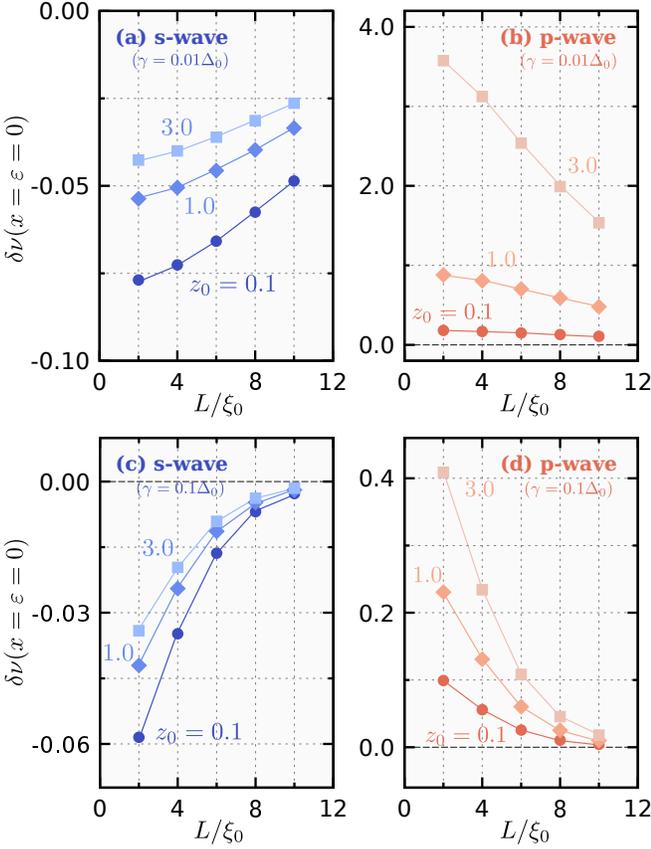}
	\caption{Junction-length dependence of the LDOS
	at $x=0$ and $\ve=0$. The results for the $s$-wave case are shown in 
	(a) and (c), where as those for $p$\,-wave are in (b) and (d). 
	The deparing ratio is fixed at $\gamma = 0.01\Delta_0$ in (a)
	and (b), and $\gamma = 0.01\Delta_0$ in (c) and (d). The
	other parameters are set to $\gamma_B = 1$ and $L-L_1 = \xi_0$. In
	$s$\,-wave cases, the correction becomes small with increasing the
	barrier potential $z_0$, whereas it becomes large with increasing
	$z_0$. The correction decreases more rapidly with increasing $L$
	when $\gamma$ is large. }
	\label{fig:ldos-Ldep-ze}
\end{figure}



The junction-length dependence of the correction at $\ve=0$ and $x=0$
(i.e., $\delta
\nu|_{x=\ve=0}$) in the $s$\,-wave VT junction is
shown in Fig.~\ref{fig:ldos-Ldep-ze}(a), where the barrier parameter 
at the interface 
is set to $z_0 = 0.1$, $1.0$, and $3.0$, and the depairing ratio is
set to $\gamma = 0.01\Delta_0$. The amplitude of the correction
$|\delta \nu|$ decreases with increasing the junction
length where the curvature of $|\delta \nu|$ as a function
of $L$ is positive. 
We have confirmed that the curvature changes at a certain length. 
In the long-junction limit (i.e., $L_1 \gg
\xi_0$), $\delta \nu|_{x=\ve=0}$ approaches to $\nu_0$ (i.e., normal
state) where the VT junction can be regarded as a pair of two T-shaped
junctions. 
%
In the $p$\,-wave junction, the amplitude of the correction $|\delta
\nu|$ decreases with increasing $L_1$ as seen in the $s$\,-wave case.
However, contrary to the $s$\,-wave case, the degree of correction decreases
with increasing $L$ more rapidly when the magnitude of 
$z_0$ is large. This behavior is
unique to the spin-triplet $p$\,-wave junction. 

The junction-length dependences with a larger depairing ratio $\gamma
= 0.1\Delta_0$ are shown in Figs.~\ref{fig:ldos-Ldep-ze}(c) and
\ref{fig:ldos-Ldep-ze}(d). Both of the $s$\,- and $p$\,-wave cases,
the amplitudes of $\delta \nu$ are smaller and decrease more rapidly
compared with the results for $\gamma = 0.01\Delta_0$. When $L=20
\xi_0$, the correction $\delta \nu$ is almost zero in all of the
cases.  Therefore, the decay length for $|\delta \nu(x,\ve=0)|$ in the
strong-proximity-effect regime 
would be mainly determined by
$\sqrt{\mathscr{D}/\gamma}$, which is consistent with the $s$\,-wave 
results with weak-proximity effect\cite{Volkov_PRL_1996}. 

\subsection{Depairing-ratio dependence}

\begin{figure}[tb]
	\centering
  \includegraphics[width=0.42\textwidth]{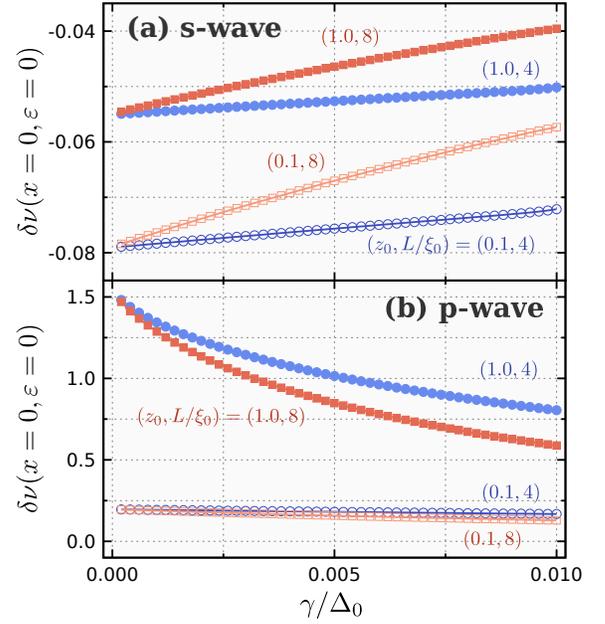}
	\caption{Depairing-ratio dependence of the LDOS
	at $x=0$ and $\ve=0$ for (a) an $s$\,-wave junction and (b) an
	$p$\,-wave junction. The interface barrier
	and the junction length are set to $z_0 = 1.0$ or $0.1$ and
	$L/\xi_0=4$ or $8$. The correction of the LDOS converges at a
	certain value regardless of the junction length, meaning that
	the decay length of $\delta \nu$ is determined by $\gamma$.  }
	\label{fig:gdep}
\end{figure}

\begin{figure}[tb]
	\centering
  \includegraphics[width=0.42\textwidth]{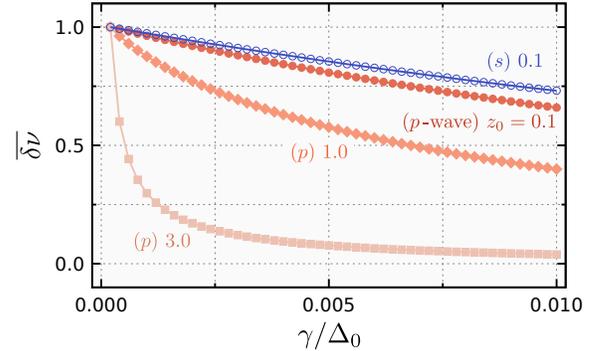}
	\caption{Depairing-ratio dependence	of the normalized correction.
	The normalized correction $\overline{\delta \nu}(\gamma)$ is given
	in Eq.~\eqref{eq:norm_cor}. The parameters are set to $\gamma_B =
	1$, $L = 8\xi_0$, and $L-L_1 = \xi_0$. For $p$\,-wave junctions, 
	the correction at the zero-energy 
	depends strongly on the interface barrier $z_0$ because 
	the large barrier potential results in the high zero-energy peak. }
	\label{fig:gdep_norm}
\end{figure}

In real samples, the deparing effects such as inelastic scatterings
are inevitably present. We lastly discuss the $\gamma$ dependence of
$\delta \nu$.  The $\gamma$ dependence of $\delta \nu|_{x=\ve=0}$
for $s$\,- and $p$\,-wave junctions are shown in
Figs.~\ref{fig:gdep}(a) and \ref{fig:gdep}(b), respectively. The
junction length and the interface barrier are fixed at $L/\xi_0=4$ or
$8$ and $z_0 = 0.1$ or $1.0$.  The corrections for $L/\xi_0=4$ and $8$
approach to a certain value even though the distance between the two S
electrodes are different. We therefore can conclude that the decay
length of $\delta \nu$ in the VT junction is determined by $\gamma$.
This behavior is consistent with that demonstrated within the
weak-proximity-effect approximation\cite{Volkov_PRL_1996,
Volkov_PRB_1997}. In the $s$\,-wave case, the slopes of $\delta \nu|_{x=0, \ve=0}$
 curves do not strongly depends on $z_0$. 

As shown in Fig.~\ref{fig:gdep}(b), the decay length of $\delta \nu$ is
determined by $\gamma$ in the $p$\,-wave junction as well. The
corrections at $\gamma = 0.001\Delta_0$ is almost independent of the
junction length, meaning which the decay length for $p$\,-wave
junction is determined by
the depairing ratio $\gamma$ as well. Contrary to the $s$\,-wave case,
however, the slopes for the $p$\,-wave junctions strongly depends on $z_0$. 

We show the $\gamma$ dependence of $\overline{\delta \nu}(\gamma)$ 
in Fig.~\ref{fig:gdep_norm}, 
where $\overline{\delta \nu}(\gamma)$ is a function
of $\gamma$ normalised by its value at $\gamma=0.001\Delta_0$; 
\begin{align}
  \overline{\delta \nu}(\gamma) = 
	\frac{\delta \nu(x=0,\ve=0;\gamma)}
	{\delta \nu(x=0,\ve=0;\gamma=0.001\Delta_0)}. 
	\label{eq:norm_cor}
\end{align}
We compare the following four cases: the $p$\,-wave junctions with $z_0 = 0.1$,
$1.0$, and $3.0$ and the $s$\,-wave junction with $z_0 = 0.1$. 
Figure~\ref{fig:gdep_norm} clearly shows that the decay length
for the $p$\,-wave junction strongly depends on $z_0$. The
$p$\,-wave result with $z_0 = 0.1$ and the $s$\,-wave results with
$z_0 = 0.1$ are not qualitatively different. 
Therefore, we conclude that the decay length for the $p$\,-wave
junction depends on the amplitude of Cooper pairs
injected by the proximity effect. 

Differing from the
N/DN/$p$\,-wave junction\cite{Tanaka_PRB_2005} where
the zero-energy LDOS at the DN/$p$\,-wave interface
diverges as $\propto 1/\sqrt{\gamma}$, the zero-energy correction $\delta
\nu (x,\ve=0)$ does not diverge even when $\gamma \to 0$ everywhere in
the DN because our system is essentially different from the system where a
$p$\,-wave SC is used as an
electrode\cite{Tanaka_PRB_2005,Tanaka_PRBR_2005}.

\section{Conclusion}
\label{sec:con}


We have theoretically studied the quasiparticle spectrum in a
junction of a diffusive normal metal where superconductors are
attached to its side surface. We have considered two types of
junctions: the T-shaped junction where one superconductor is attached
to the diffusive normal metal and the Volkov-Takayanagi junction where
two superconductors are attached to it.  In the T-shaped junction,
when the superconductor is spin-singlet $s$-wave, the local density of
states which can be measured by scanning tunneling
spectroscopy (STS) measurements has a dip structure which is
consistent with the standard proximity effect.  On the other hand, in
the spin-triplet $p$-wave case, there is a zero-energy peak in the
local density of states due to the anomalous proximity effect by
odd-frequency pairing.  The amplitude of the correction in the local
density of states is strongly depends on the interface barrier. In the
$p$\,-wave case, in particular, the larger barrier results in the
larger density of states at the zero energy.

In the Volkov-Takayanagi junction, the phase difference between
the two superconductors affects significantly on the energy spectrum. In the $s$\,-wave junction without
a phase difference, the low-energy dip appears at the center of
the junction.  On the contrary, when the phase
difference is $\pi$, such a low-energy dip vanishes and the local
density of state at the center becomes one in the normal state because
of the destructive interference of Cooper pairs. 
When spin-triplet $p$\,-wave superconductors are employed instead of
spin-singlet $s$\,-wave superconductors, the zero-energy resonant
states appear.  When there is no phase difference, the zero-energy
peak spreads spatially between the two superconductors, whereas the
peak vanishes at the center of the junction when the phases differ by
$\pi$.  

We have also studied the characteristic length scale of the
phase coherence.  We have shown that, in both of the $s$\,-wave and
$p$\,-wave cases, the decay length of the zero-energy state is mainly
characterized by the depairing ratio $\gamma$ by, for example,
inelastic scatterings.  We have
demonstrated that the decay length is not simply determined by
$\gamma$ for spin-triplet $p$\,-wave junctions.  The decay length for
a $p$\,-wave junction depends also on the quality of the interface
because the strength of the resonance depends strongly on the
interface barrier potential.

\begin{acknowledgments}
The authors would like to thank T.~Yokoyama and S.~Tamura 
for useful discussions. 
This work was supported by Grants-in-Aid from JSPS for Scientific
Research on Innovative Areas ``Topological Materials Science''
(KAKENHI Grant Numbers JP15H05851, JP15H05852, JP15H05853 and JP15K21717), 
Scientific Research (B) (KAKENHI Grant Number JP18H01176), 
Japan-RFBR Bilateral Joint Research Projects/Seminars number 19-52-50026,  
JSPS Core-to-Core Program (A. Advanced Research Networks).
A.~A.~G. acknowledges supports by the European Union
H2020-WIDESPREAD-05-2017-Twinning project ``SPINTECH'' under grant
agreement Nr.~810144. 
\end{acknowledgments}

\appendix*

\section{Additional symmetry of the Usadel equation}
In the quasiclassical formalism, the anomalous Green's functions $f$
and $\ut{f}$ are related by several symmetry relations. In a diffusive system
(i.e., Usadel formalism), the Green's functions can have additional
symmetry compared with the ballistic case. 

\subsection{General symmetry} 
The Usadel equation
for the retarded and advanced component is given by 
\begin{align}
  & D \bs{\nabla}_{\bs{r}} 
  \big( \check{g}_o^X \bs{\nabla}_{\bs{r}} \check{g}^X_o \big)
	+ i \big[ \check{H}^X_o, \check{g}^X_o \big]_- = 0, 
	\label{eq:ap:Usa}
	\\[4mm]
	& \check{H}^X_o
	= \left[ \begin{array}{cc}
	  \ve^X \hat{\sigma}_0 & 
		\hat{\Delta}(\bs{r},\ve) \\[2mm]
	  \hat{\Delta}^* (\bs{r},-\ve)& 
		-\ve^X \hat{\sigma}_0 \\
	\end{array} \right], 
	\\[2mm]
	&\check{g}^X_o(\bs{r},\ve)
	 = \left[ \begin{array}{cc}
	       \hat{g}^X(\bs{r},\ve) & 
		     \hat{f}^X(\bs{r},\ve) \\[2mm]
	  -\ut{\hat{f}}^X(\bs{r},\ve) & 
	  -\ut{\hat{g}}^X(\bs{r},\ve) \\
	\end{array} \right], 
\end{align}
where $\check{g}^X_o$ with $X = R$ $(A)$ means retarded (advanced)
Green's function.  Assuming the single-component pair potential (i.e.,
either of the even-frequency spin-singlet or odd-frequency
spin-triplet SCs), the matrix $\check{H}$ becomes 
\begin{align}
	 \check{H}^X_o
	&= \left[ \begin{array}{cc}
	  \ve^X \hat{\sigma}_0 & 
		     \Delta_\mu   (\bs{r}) i \hat{\sigma}_\mu   \hat{\sigma}_2 \\[2mm]
	  s_\ve\Delta_\mu^* (\bs{r}) i \hat{\sigma}_\mu^* \hat{\sigma}_2 & 
		-\ve^X \hat{\sigma}_0 \\
	\end{array} \right]
	\\[2mm]
	&= 
	\left[ \begin{array}{cc}
	  \ve^X \hat{\sigma}_0 & 
		\Delta_\mu   (\bs{r})(i\hat{\sigma}_\mu \hat{\sigma}_2  ) \\[2mm]
	  \Delta_\mu^* (\bs{r})(i\hat{\sigma}_2   \hat{\sigma}_\mu) & 
		-\ve^X \hat{\sigma}_0 \\
	\end{array} \right]. 
\end{align}
where $\Delta_\mu(\bs{r}) \in \mathbb{C}$ is the scalar pair potential
with $\mu \in [1,3]$.
The factor $s_\ve$, which is defined as $s_\ve = +1$ $(-1)$ for
even-frequency (odd-frequency) SCs, stems from the frequency symmetry
of the pair potential. We have used $s_\ve s_\mu = -1$ (i.e., Pauli
rule) and $\hat{\sigma}_\mu^* \hat{\sigma}_2 = -s_\mu \hat{\sigma}_2
\hat{\sigma}_\mu$. In this case, it is convenient to parametrise the
spin structure of the Green's function as following: 
\begin{align}
	& \check{g}^X_o
	= 
	\left[ \begin{array}{cc}
	       g ^X  \hat{\sigma}_0 & 
		     f ^X_\mu(i\hat{\sigma}_\mu \hat{\sigma}_2  ) \\[2mm]
	  -\ut{f}^X_\mu(i\hat{\sigma}_2   \hat{\sigma}_\mu) & 
		-   {g}^X  \hat{\sigma}_0 \\
	\end{array} \right]. 
\end{align}
We can simplify the Usadel equation by the unitary transform. We first
define the unitary matrix:
$
  [\check{U}_1]^{-1}
	=
	\mathrm{diag} \left[ \hat{\sigma}_0, -i\hat{\sigma}_2\hat{\sigma}_\mu \right]
$. 
Multiplying $\check{U}_1$ and $\check{U}_1^{-1}$ from the left and
right side of the Usadel equation \eqref{eq:ap:Usa}, we have the
simplified Usadel equation: 
\begin{align}
  & D \bs{\nabla}_{\bs{r}} 
  \big( \check{g}^X \bs{\nabla}_{\bs{r}} \check{g}^X \big)
	+ i \big[ \check{H}^X, \check{g}^X \big]_- = 0, 
	\\[4mm]
	& \check{g}^X(\bs{r},\ve)
	= \left[ \begin{array}{rr}
	       g ^X & 
		     f ^X \\[2mm]
	   \ut{f}^X & 
	  -    g ^X \\
	\end{array} \right]
	\otimes \hat{\sigma}_0, 
	\hspace{6mm}
	\\[2mm]
	& \check{H}^X{(\bs{r},\ve)}
	= \left[ \begin{array}{cc}
	  \ve^X & 
		\Delta  (\bs{r}) \\[2mm]
	  \Delta^*(\bs{r})& 
		-\ve^X \\
	\end{array} \right] \otimes \hat{\sigma}_0, 
\end{align}
where we redefine the Greens functions and the matrix $\check{H}^X$ as
following: 
$\check{g}^X{(\bs{r},\ve)} = 
\check{U}_1 \check{g}^X_0{(\bs{r},\ve)} \check{U}_1^{-1}$ and 
$\check{H}^X{(\bs{r},\ve)} = 
\check{U}_1 \check{H}^X_0{(\bs{r},\ve)}
\check{U}_1^{-1}$, and the subscript $\mu$ is omitted. 

The matrix $\check{H}^X(x,\ve)$ satisfies several
symmetric relations. Hereafter, we consider the one-dimensional
system. Using the Pauli matrices in the particle-hole space, we can
express the matrix $\check{H}^X(x,\ve)$ with a simpler form: 
\begin{align}
  \check{H}^X(x,\ve) 
	= \check{\tau}_3 \ve^X
	+i\check{\tau}_2 \Delta_{\mathrm{R}}(x) 
	+i\check{\tau}_1 \Delta_{\mathrm{I}}(x), 
\end{align}
where $\Delta_{\mathrm{R}(\mathrm{I})} \in \mathbb{R}$ is the real
(imaginary) part of the pair potential. 
The first symmetry is given by 
\begin{align}
  & \check{H}^R(x,\ve) 
	= 
	- \check{\tau}_1
    \left[ \check{H}^{A}(x,\ve) \right]^*
	  \check{\tau}_1, 
	\\[4mm]
  & {g}^R(x,\ve) = -\big[    {g}^A(x,\ve) \big]^*, \\[0mm]
  & {f}^R(x,\ve) =  \big[ \ut{f}^A(x,\ve) \big]^*, 
\end{align}
where we have used $\ve^R = [\ve^A]^*$. 
The relations above connect the retarded and advanced Green's
functions. The second symmetry is given by 
\begin{align}
  & \check{H}^X(x,\ve) 
	= 
	\check{\tau}_1 
  \left[ \check{H}^{X}(x,-\ve) \right]^*
	\check{\tau}_1, 
	\\[4mm]
  & {g}^X(x,\ve) = \big[    {g}^X(x,-\ve) \big]^*, \\[0mm]
  & {f}^X(x,\ve) = \big[ \ut{f}^X(x,-\ve) \big]^*, 
\end{align}
The third symmetry is given by 
\begin{align}
   \check{H}^X(x,\ve) 
	&= -
	\check{U}_\phi
  \left[ \check{H}^{X}(x,\ve) \right]
	\check{U}_\phi, 
	\label{eq:sym3}
	\\[2mm]
   \check{U}_\phi
	&= \left[ \begin{array}{cc}
	  & 
		e^{ i \phi} \\[2mm]
		e^{-i \phi} &
		\\
	\end{array} \right], 
\end{align}
where $\phi(x)$ is the local phase defined as  $\phi = \mathrm{arctan}
\left( \Delta_{\mathrm{I}} /  \Delta_{\mathrm{R}} \right)$. 
We can reduce the following relations from Eq.~\eqref{eq:sym3}: 
\begin{align}
  {f}^X(x,\ve) e^{-i \phi(x)} 
	= - \ut{f}^X(x,\ve) e^{ i \phi(x)}. 
	\label{eq:sym6}
\end{align}
When the pair potential is a real function, we can parametrise the
Green's function as 
\begin{align}
	& \check{g}^X(\bs{r},\ve)
	= \left[ \begin{array}{rr}
	   g^X & 
		 f^X \\[2mm]
	  -f^X & 
	  -g^X \\
	\end{array} \right]
	\otimes \hat{\sigma}_0. 
\end{align}

\subsection{Symmetry in Josephson(-ish) junctions}
In Josephson(-ish) junctions, the Green's functions have additional
symmetry.  In this paper, we refer to the junctions
in which the relation $\phi(x) = - \phi(x)$ is satisfied as the
Josephson-ish junctions (e.g., Volkov-Takayanagi junctions). 
In other words, the real and imaginary parts of the pair potential
are even and odd function of $x$: 
\begin{align}
  &\Delta_{\mathrm{R}}(x) =  \Delta_{\mathrm{R}}(-x), \\[2mm]
  &\Delta_{\mathrm{I}}(x) = -\Delta_{\mathrm{I}}(-x). 
	\label{eq:Delta}
\end{align}
In this case, the matrix $\check{H}^X(x,\ve)$ and the Green's functions 
satisfy the symmetry relations related to the real space: 
\begin{align}
  & \check{H}^X(x,\ve) 
	= -
	\check{\tau}_1
  \check{H}^{X}(-x,\ve)
	\check{\tau}_1, 
	\label{eq:sym4}
	\\[2mm]
  & {g}^X(x,\ve) =     {g}^X(-x,\ve), \\
  & {f}^X(x,\ve) = -\ut{f}^X(-x,\ve). 
	\label{eq:sym5}
\end{align}
Combining Eqs.~\eqref{eq:sym6} and \eqref{eq:sym5}, we have 
\begin{align}
    {f}^X(x,\ve)e^{-i \phi(x)} 
	= {f}^X(-x,\ve) e^{ i \phi(x)}. 
	\label{eq:sym7}
\end{align}
In particular, the relation above can further be reduced when the
phase difference is either $\delta \Phi = 0$ or $\pi$: 
\begin{align}
  \left\{ \begin{array}{cl}
    {f}^X(x,\ve) = +{f}^X(-x,\ve) & \text{~~for~~ $\delta \Phi = 0$, }\\[3mm]
    {f}^X(x,\ve) = -{f}^X(-x,\ve) & \text{~~for~~ $\delta \Phi = \pi$. }\\
	\end{array} \right.
	\label{eq:sym7}
\end{align}

\bibliography{tsc05}

\end{document}